\documentclass[showpacs,preprintnumbers,amsmath,amssymb,nofootinbib]{revtex4}
          \usepackage{graphicx}
          
\begin{document}

          \title{ Bose-Einstein Condensation Picture of Superconductivity in High
                 and in Low-Temperature Superconductors  (Dilute Metals).}
          \author{V. N. Bogomolov}
          \affiliation{A. F. Ioffe Physical \& Technical Institute,\\
         Russian Academy of Science,\\
         194021 St. Petersburg, Russia} \email{V.N.Bogomolov@inbox.ru}
         \date{\today}
         \begin{abstract}
              Structures and parameters of some high and low temperature superconductors (HTSC, LTSC)
              are considered basing on the alternative estimate of the $\textrm{O}^{2-} \textrm{ion}$ radius magnitude $(0.5-0.6)$\!\AA.
              Phase transitions into the superconducting state are considered as the Bose-Einstein condensation (BEC).
              The super HTSC with $T_{c} = 371\,\textrm{K}\; (\textrm{YBa}_{2}\textrm{Cu}_{3}\textrm{Se}_{7})$
              and $T_{c} \sim 400\,\textrm{K}\;(\textrm{Ag}_{2}(\textrm{Ag}_{3}\textrm{Pb}_{2}\textrm{H}_{2}\textrm{O}_{6})),$
              and LTSC with $T_{c}\sim0.3\,\textrm{K}\;
              (\textrm{SrNb}_{x}\textrm{Ti}_{(1-x)}\textrm{O}_{3})$ are shown to be of the BEC type. Instability of the structure of the first one results
             from higher magnitude of the $\textrm{Se}^{2-} \textrm{ion}$ radius in comparison with the $\textrm{O}^{2-}$\!radius. The second one forms quasi
              one-dimensional structures and is rather inpractical. The electron density and the effective mass are estimated
              for some stoichiometric and non-stoichiometric (nano-composite) high temperature superconductors, which have
              some peculiar features. Large effective masses can indicate existance of polarons (bipolarons) in  such
              systems. Some new superconductors \!$(\textrm{Mg}_{x}\textrm{WO}_{3}).$
            \end{abstract}
            \maketitle
            \bigskip
                 The problem of HTSC physical mechanism remains open up to now. Fabrication of superconductors with $T_{c} > 300\,\textrm{K}$
                 or at least confirmation of information on unexpected synthesis of such materials somewhere would
                 be extremely important as for physics so for electronics and electro-and magneto-energetics.
                 It is quite possible that the first communication on discovery of a material with $T_{c}=371\,\textrm{K}$ \cite{bib1}
                 was not confirmed because of instability of materials and irreproducibility of synthesis.
                 Oxygen in $\textrm{YBa}_{2}\textrm{Cu}_{3}\textrm{O}_{7}$ was substituted with selenium in
                 $\textrm{YBa}_{2}\textrm{Cu}_{3}\textrm{Se}_{7}$ in this case.

             1). One of the main reasons of the current situation is a use of the traditional magnitude of
             $1.4$\!\!\AA \;  for the oxygen ion $\textrm{O}^{2-}$\! radius instead of the realistic in our opinion
             value of $0.5 - 0.6$\AA \;\cite{bib2}.
             It is very important also the closeness of the radii $r_{\textrm{O}^{2-}}$ and $r_{\textrm{O}}$ and relation $r_{\textrm{Me}}
             >>r_{\textrm{Me}^{2+}}.$
             A cardinal transformation of the oxide electronic structure due to this conjecture is depicted
             schematically in $\textrm{Fig}. \,1 \;a),\, b).$ An instability of the structure occures due to a substitution
             of $\textrm{O}^{2-}$ by $\textrm{Se}^{2-}$ ions with radius, which is twice as large as for the former. Structures of the
             stoichiometric $\textrm{YBa}_{2}\textrm{Cu}_{3}\textrm{Se}_{7}$ and nonstoichiometric
             $(\textrm{Ag}_{2}(\textrm{Ag}_{3}\textrm{Pb}_{2}\textrm{H}_{2}\textrm{O}_{6}))$
           with  $T_{c} > 400\,\textrm{K}$ \cite{bib3}
             are presented in Fig.\! 2 and Fig.\! 3. A sectional view of the $\textrm{SrTiO}_{3}$ unit cell
             is depicted in Fig.\! 4.

                  2). The second reason of the obstacle in understanding of HTSC is due to a use of the
                  stationary ion radii as it is accepted for classical ionic compounds $(\textrm{Fig}.\,1a)$ instead of
                  the quantum superposition of the ground and ionized (excited) states (Fig.1b). Electron
                  properties of the oxide are determined by tunneling between these states \footnote{\footnotesize{The quantum superposition of the ground and excited
                  states (the atom remains neutral) takes place in atoms of the
                 noble gases condensates and of the metals like Pd. Their
                 metallization occurs due to tunneling over the excited states [4].}}). Their occupancy
                  is inversely proportional to $r^{3},$ where \textit{r} is the given orbital radius.  The metal atom ground
                  state orbital occupancy is small $z_{Me} < 1$ (Fig.\! 1\! b) that corresponds to the diluted metal
                  limit ("chemical" dilution; a role of  the ionicity  and adiabatic effects is high).
                  The electron density (taking into account the metal atom ground state occupancy $z_{\textrm{Me}} < 1$)
                  can be not large enough for a transition of the system into the metal state and the electron
                  pairs of the divalent atoms will survive, though their binding energy  $\Delta_{0}$  drops by a factor
                  of $\varepsilon^{*}$ down to the $\Delta_{c}$ value. The superconducting phase transition arises in this case due to
                  the Bose-Einstein condensation (BEC) of the pairs not due to the Bardeen-Cooper-Schrieffer (BCS)
                  mechanism. Notice that no pairs exist in univalent atoms. The BCS theory $(\varepsilon^{*}_{\textrm{Met}})$ does not
                  explain superhigh  transition temperatures and superlow (or vanishing) transition temperatures
                  in univalent metals at standard conditions. The atomic electron pair binding energy
                  (it equals $\Delta_{0}= 5.2\,\textrm{eV}$ for Ba and  $\Delta_{0}=10.1\,\textrm{eV}$ for Hg)
                  decreases by a factor
                  of $\varepsilon^{*} (\Delta_{c}= {\frac{\Delta_{0}}{\varepsilon^{*}}}),$ when the atoms are moved together
                  (here $\varepsilon^{*}$ is the effective
                  dielectric permeability of the medium containing $\textrm{Me}^{2+}$ ions.) No such "relict"
                  electron pairs exist in univalent metals. Properties of the medium change abruptly
                  at metallization $(\varepsilon ^{*}<< \varepsilon^{*}_{_{\textrm{Met}}})$ and atomic electron pairs give place (or transmute)
                  to the Cooper ones. The "chemical" dilution allows us to obtain stable systems at
                  the state near to the metallization threshold \cite{bib5}. The energy equality condition
                  for electronic energies in the phases at the transition point
                  $\frac{2.87(\frac{h}{2\pi})^{2})n^{5/3}}{m^{*}} \sim \frac{\Delta_{c}n}{2}$  gives $T_{cmax} \sim
                  \frac{\Delta_{c}}{5.5k}$  \cite{bib5}.
                  In $\textrm{YBa}_{2}\textrm{Cu}_{3}\textrm{O}_{7},$
                  $T_{c} = 90\,\textrm{K};$ \; $r_{\textrm{Ba}} = 2.19$\!\!\AA; \; $\varepsilon^{*} \sim 25;$ \;
                  $m^{*} \sim 5m_{e};$
                  \; $\Delta_{c}\sim 5.2/25 = 0.21 \,\textrm{eV} \sim 5.5kT_{cmax}$  \cite{bib5}.
                  That is why the real $T_{c} = 90\,\textrm{K}$ is significantly less, than possible $T_{cmax} \sim 495\,\textrm{K}$
                  and the transition has features of the Bose-Einstein condensation of hard electron
                  pairs. An increase of the pair density $n_{2}$ will lead to a rise of $T_{c}$, but only up
                  to $495\,\textrm{K}$, i.e. the system metallization temperature. Then $\varepsilon^{*}$ changes abruptly,
                  $\Delta_{c}$ drops, and the pairs transmute into the Cooper ones, while the transition acquires
                  BCS character being accompanied by depairing \cite{bib5}.

               3).  A density of the "active" electrons can be estimated assuming that dilution is determined by
               occupancies of the metal atom orbitals, which depend on the acceptor properties of oxigen (or selen).
               The three-dimensional superconductivity occurs in result of tunneling over the electron quantum
               states of the metal atoms. An irregular network ("cobweb") of paths and rings \cite{bib4,bib6,bib7}
               gives way
               probably to regular $3d$ lattice of superconducting threads ("gossamer") \cite{bib8}.

              The metal and oxygen orbital occupancies are inversely proportional to the orbital volume
              $z_{\textrm{Me}}/z_{\textrm{O}} \sim (r_{\textrm{O}}/r_{\textrm{Me}})^{3}$ \cite{bib5}.
              For the pair $\textrm{Se}^{2-}-- \;\textrm{O}^{2-}$
              we have $z_{\textrm{MeSe}} \sim 8z_{\textrm{MeO}}.$ This gives for
              the BEC model the transition temperature increase of $8^{\frac{2}{3}} = 4$times, i. e.
              up to $360\,\textrm{K}$ that
              complies with the data \cite{bib1}. Such electron pairs density corresponds to the composition of
              $\textrm{YBa}_{2}\textrm{Cu}_{3}\textrm{O}_{7/8}$ (a metal alloy mixed with oxides).
              Replacement of Ba atoms with Hg ones
              $(\Delta_{0}=10.1 \,\textrm{eV};\; r = 1.45$\!\!\AA) leads to increase of the pair density by a factor
              of $(2.19/1.45)^{3} \sim 3.4$ and of $T_{c}$ up $T_{p}\sim 200\,\textrm{K}.$ The BEC threshold increases
              up to $T_{max} \sim 1000\,\textrm{K}.$

             4). One oxygen atom falls at one metal atom in $\textrm{YBa}_{2}\textrm{Cu}_{3}\textrm{O}_{7}.$
             A number of electron pairs in the unit
             cell equals $2.5;$ each pair occupies the volume of $69.4$\!\AA\!\!\!$^{3}.$ If one takes into account the Ba
             atom orbital occupancy $z_{\textrm{Me}} \sim z_{\textrm{O}}\, (0.5/2.19)^{3}  = z_{0}/84,$ the "active"
             electron pairs density takes the value of $n_{2} \sim 1.7 \cdot 10^{20} \textrm{cm}^{-3}.$
              In the BEC case $T_{c}= 92\,\textrm{K}$ for $m^{*} \sim 5m_{e}.$ The medium dielectric constant  $\varepsilon^{*}\sim
              20.$

             5).   In the case of $\textrm{SrTiO}_{3},$ the atomic value  $\Delta_{0\textrm{Sr}} = 5.7\,\textrm{eV}$
             becomes equal to
              $\Delta_{c} \sim 1.4\cdot 10^{-4}\,\textrm{eV}$ at $\varepsilon^{*} \sim 4\cdot 10^{4}$ \cite{bib9}
             and $T_{cmax} =  \frac{\Delta_{c}}{5.5\textrm{k}} \sim 0.25\,\textrm{K}.$ An estimate of the intrinsic electron
             density  basing on the orbital
             occupancy for the Sr atoms $( n \sim r^{-3} )$ gives $n \sim 10^{20}\textrm{cm}^{-3}.$ This estimate is too rough since
             the Sr~ - ~O distance is distinctly larger, than $r_{\textrm{Sr}} + r_{0}$ ( Fig. 4 ). Doping by Nb partly blocks
             the oxigen valence and increases the free electron pairs density given by Sr atoms
             $(\textrm{SrNb}_{x}\textrm{Ti}_{(1-x)}\textrm{O}_{3})$ \cite{bib10}. $T_{c}$ is proportional
             to $n^{2/3}$ up to $T_{cmax} = 0.28\,\textrm{K}$ \cite{bib10}.
             The temperature $T_{c}$
             falls abruptly after $T_{cmax} (n > 1 \cdot 10^{20} \textrm{cm}^{-3});$ metallization occurs and BCS superconductivity
             establishes \cite{bib11}. The function $T_{c} \sim n$ \cite{bib10} is asymmetric and corresponds to the plot in Fig. 5.
             The electron effective mass equals about $1200 \cdot m_{e}.$

              6).  Atomic cores of the electron pairs participate in forming of the medium electronic
              properties $(\varepsilon^{*}, m^{*})$ of stoichiometric compounds. The metal atoms are
              diluted by oxigen ones.
              In nonstoichiometric (nanocomposite) HTSC compounds atoms and their electron pairs have contact
              only with atoms of the matrix (Fig. 3). The metal atoms are diluted by matrix. The electron
              pairs density and the binding energies in diatomic molecules are known for the "physical" dilution
              of metals \mbox {in
              \!\!$\textrm{Ag}_{2}(\textrm{Ag}_{3}\textrm{Pb}_{2}\textrm{H}_{2}\textrm{O}_{6})\quad
               (T_{c}\sim 400 \,\textrm{K}; \; m^{*}\sim 7.5 \textrm{m}_{e})$\! \cite{bib3},}
                \mbox{in $(\textrm{Na}_{2})_{0.02}\textrm{NH}_{3} \quad (T_{c} \sim 200\,\textrm{K};$
              $m^{*} \sim 5 \textrm{m}_{e})$ \cite{bib12} \quad and} \mbox{in $(\textrm{Na}_{2})_{0.025}\textrm{WO}_{3}\quad
              (T_{c} \sim 91\,\textrm{K}; \quad m^{*} \sim 10 m_{e})$ \cite{bib13}.}
              This allows us to estimate the electron
              effective mass $m^{*}$ basing on the BEC magnitude of $T_{c}.$ There are $3d$ lattices of $\textrm{Ag}_{2}$
               molecules chains
              or $3d$ networks of irregular $\textrm{Na}_{2}$ chains in these compounds. These chains seemingly weakly interact
              with the matrix. In result, the pair binding energy  $\Delta_{c}$ is almost independent on the matrix
              $\varepsilon^{*}.$
              In the all of three cases, the pair binding energy can be estimated as  $\Delta_{c} \sim 5.5 kT_{cmax}$ \cite{bib5}.
              The intermediate case between BEC and BCS occurs. The electron effective masses in the all
              of four systems \cite{bib1,bib3,bib12,bib13} are very near and can stem from the polaron (bipolaron) effect \cite{bib14}.

                     7).  Parameters of some HTSC are given in the Table.
        \begin{table}[h]

         \begin{tabular}{|l|c|c|c|c|c|} 
         \hline
         & $ \textrm{YBa}_{2}\textrm{Cu}_{3}\textrm{O}_{7}$ & $\textrm{YBa}_{2}\textrm{Cu}_{3}\textrm{Se}_{7}$
        & $\textrm{Ag}_{2}(\textrm{Ag}_{3}\textrm{Pb}_{2}\textrm{H}_{2}\textrm{O}_{6})$
        & $(\textrm{Na}_{2})_{0.025}\textrm{WO}_{3}$ &  $(\textrm{Na}_{2})_{0.02}\textrm{NH}_{3}$ \\ \hline
         $T_{c} \textrm{K} $ & $ \sim 90 $  &  $ 371 $ & $ \sim 400 $ & $ \sim 90  $ & $ \sim 180 $\\ \hline
         $T_{cmax}{K} $ & $ \sim 500 $ & $ \sim 500 $ & $ \sim 500 $ & $ \sim 100 $ & $ \sim 200 $\\ \hline
         $n_{2} {\textrm{cm}}^{-3} $ & $ 1.7\cdot10^{20}  $ & $ 13.6\cdot10^{20}$ &
         $25.6\cdot10^{20} $ & $ 4.74\cdot10^{20}  $ & $ 4.71\cdot10^{20} $\\ \hline
         $\Delta_{c} {eV}$ & $ >0.21 $ & $ \sim 0.173 $ & $ \sim 0.187 $ & $ \sim 0.043 $ & $ \sim 0.09 $\\ \hline
         $m^{*}/m_{e} $ & $\sim 5 $ & $ \sim 5 $ & $ \sim 7.5 $ & $ \sim 10 $ & $ \sim 5 $\\ \hline
         $\varepsilon^{*}$ & $25(?)(90\,\textrm{K}) $ & $ \sim 25(?) $ & $ \sim (30-40)(?) $ & $ \sim 50(100\,\textrm{K}) $ &
         $ 25.4(200\,{\textrm{K}}) $\\ \hline

        \end{tabular}
        \end{table}

              8). Conclusion.

            "Physical" or "chemical" dilution of metals allows us to synthesize systems, intermediate between
            insulator and metal.        A possibility to explain superconductivity of some systems as a result
            of the Bose-Einstein condensation of the electron pairs belonging to diatomic molecules formed by
            univalent atoms or the electron pairs of divalent atoms was shown above. However, an increase of $T_{c}$
            can be accompanied by instability or a quasi-one-dimensional structure can appear. Practically
            important superconductors must be $3\textrm{-dimensional},$ have small  $\varepsilon^{*},$ large
            $\Delta_{c},$
            have maximum density of homogeneously distributed electrons (before metallization).
            Synthezis of such materials must obey
            to a few physico-chemical conditions simultaneously.  A directed physico-chemical "constructing"
            of superconductors can be most easily done in the case of nonstoichiometric systems - nanocomposites
            (a matrix filled with properly chosen atoms).  $\textrm{Mg}_{x}\textrm{WO}_{3}\; (\Delta_{0}\sim 7.6 \,\,\textrm{eV};\quad T_{cmax}\sim 600\,\textrm{K}$ at $x \sim (0.3-0.4)$)
            can be considered as one of examples of such materials, which is a nanocomposite unlike to $\textrm{MgB}_{2}.$
            In the latter, Mg is diluted "chemically", the compound is a metal with the BCS type superconductivity.
            Several equilibrium Mg atom sublattices with $T_{cx} < 600\,\textrm{K}$ in the cavity lattice of $\textrm{WO}_{3}$ are possible.

                The given here consideration show that synthesis of superconductors with $T_{c} > 300\,\textrm{K}$ is really possible.

        \begin{figure*}[tbp]
        \includegraphics[scale=0.8]{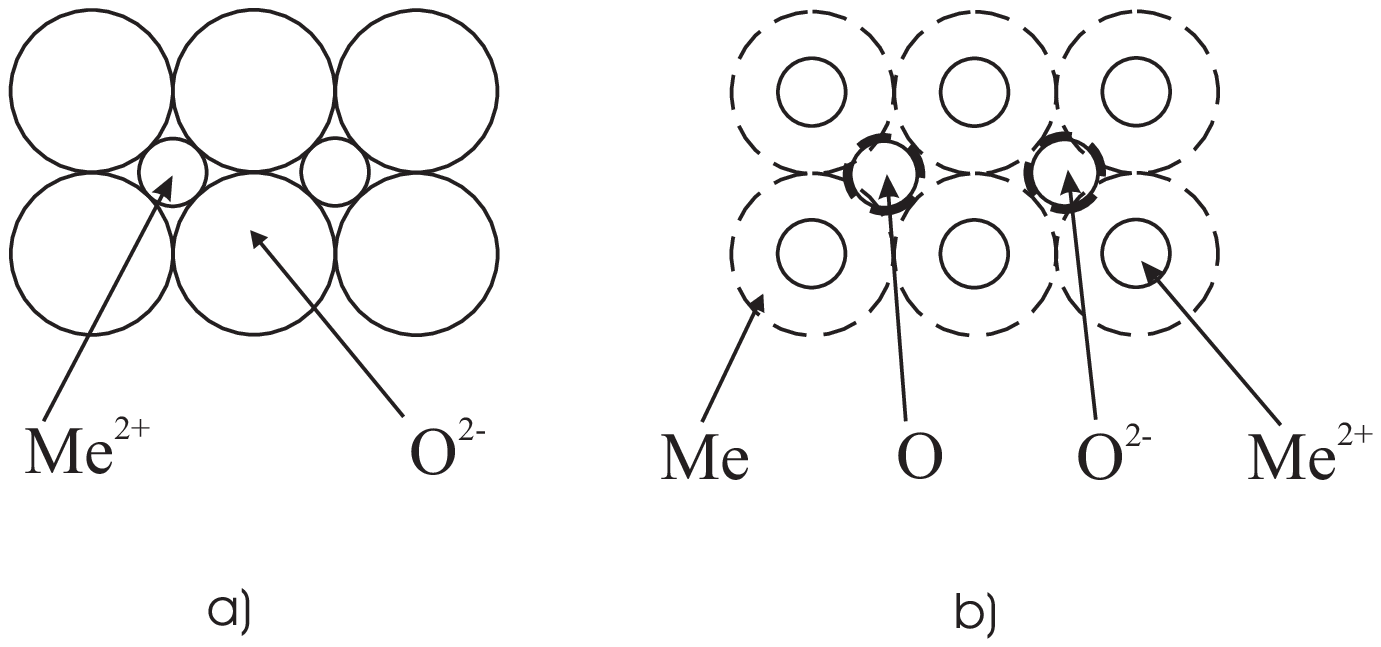}
        \caption{ Schematic representation of some possible structures of oxides:\\
         $a)$  the $\textrm{O}^{2-}$ ion radius is equal to $r_{\textrm{O}^{2-}} \sim 1.4$\!\AA;\\
         $b)$   $r_{\textrm{O}^{2-}} \sim r_{\textrm{O}} \sim 0.5$\!\!\AA (Dilute metal).}
        \end{figure*}

\begin{figure*}[tbp]
        \includegraphics[scale=0.8]{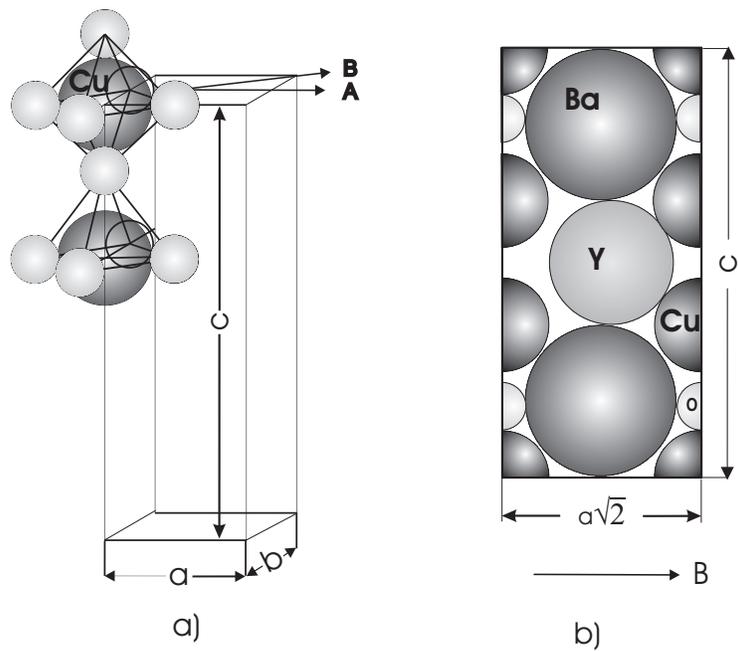}
        \caption{$a)$ the unit cell of $\textrm{YBa}_{2}\textrm{Cu}_{3}\textrm{O}_{7},$\\
         $b)$  the cross section of the unit cell along B-direction.}
        \end{figure*}

\begin{figure*}[tbp]
        \includegraphics[scale=0.8]{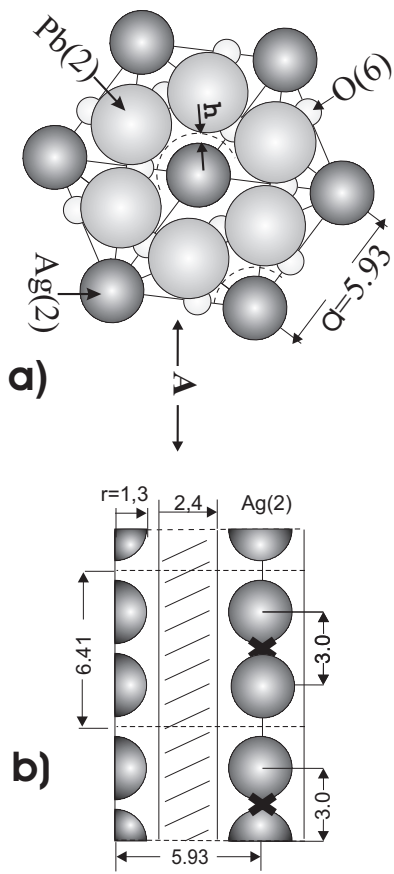}
        \caption{Packing  of atoms in $\textrm{Ag}_{2}(\textrm{Ag}_{3}\textrm{Pb}_{2}\textrm{H}_{2}\textrm{O}_{6}):$ \\
        $a)$- a view along c-axis;\\  $b)$- 3D -lattice of quasionedimentional chains of $\textrm{Ag}_{2}$ molecules.}
        \end{figure*}

\begin{figure*}[tbp]
        \includegraphics[scale=0.8]{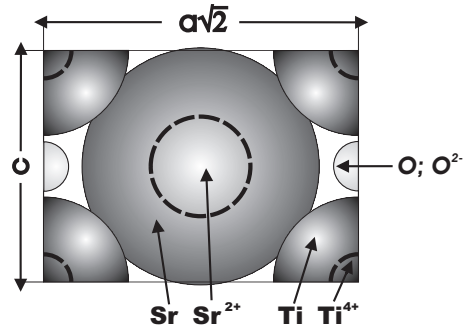}
        \caption{A sectional view of the $\textrm{SrTiO}_{3}$ unit cell.}
        \end{figure*}

\begin{figure*}[tbp]
        \includegraphics[scale=0.8]{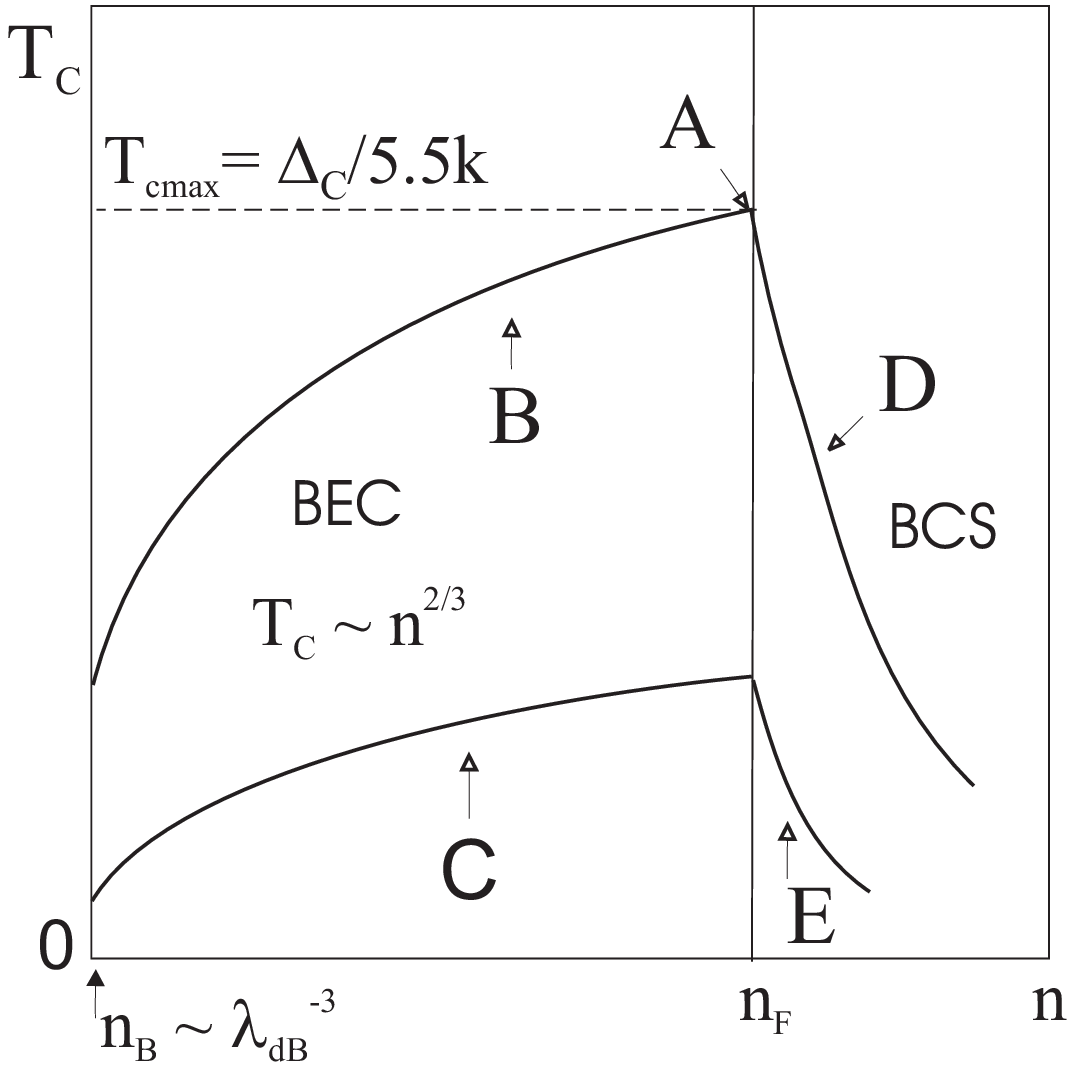}
        \caption{Schematic representation  of\! \!$T_{c}$\! dependance on the electron concentration for:\\
        \textbf{A} \textemdash  \; $\textrm{YBa}_{2}\textrm{Cu}_{3}\textrm{Se}_{7};$
        $\textrm{Ag}_{2}(\textrm{Ag}_{3}\textrm{Pb}_{2}\textrm{H}_{2}\textrm{O}_{6});$
         $(\textrm{Na}_{2})_{0.025}\textrm{WO}_{3};$     $(\textrm{Na}_{2})_{0.02}\textrm{NH}_{3};$
         $\textrm{SrNb}_{y}\textrm{Ti}_{(1-y)}\textrm{O}_{3};$   $\textrm{Mg}_{y}\textrm{WO}_{3}.$
       \\\qquad \textbf{ B}\; \textemdash \; $\textrm{YBa}_{2}\textrm{Cu}_{3}\textrm{O}_{7};$ $\textrm{Mg}_{x}\textrm{WO}_{3}.$
       \\\qquad  \textbf{C}  \; \textemdash  \; $\textrm{SrNb}_{x}\textrm{Ti}_{(1-x)}\textrm{O}_{3}.$
       \\\qquad \textbf{ D}  \; \textemdash  \; $ \textrm{MgB}_{2}.$
       \\\qquad \textbf{E}  \; \textemdash  \; $\textrm{SrNb}_{z}\textrm{Ti}_{(1-z)}\textrm{O}_{3}.$ $(x<y<z)$
       \\\qquad
       \textbf{$\lambda_{\textrm{dB}}$}\textemdash \; the de Broglie wavelength.}
        \end{figure*}

        \end{document}